# Reversible Data Hiding in Encrypted Text Using Paillier Cryptosystem

Asad Malik *, Aeyan Ashraf[†], Hanzhou Wu[‡] and Minoru Kuribayashi [§]
* Department of Computer Science, Aligarh Muslim University, Aligarh 202002, India
[†] Department of Computer Engineering, Aligarh Muslim University, Aligarh 202002, India
[‡] School of Communication and Information Engineering, Shanghai University, Shanghai 200444, China
[§] Graduate School of Natural Science and Technology, Okayama University, Okayama 700-8530, Japan
Email: *amalik.cs@amu.ac.in, [†]gj3323@myamu.ac.in, [‡]h.wu.phd@ieee.org, [§]kminoru@okayama-u.ac.jp

*Abstract*—Reversible Data Hiding in Encrypted Domain (RD-HED) is an innovative method that can keep cover information secret and allows the data hider to insert additional information into it. This article presents a novel data hiding technique in an encrypted text called Reversible Data Hiding in Encrypted Text (RDHET). Initially, the original text is converted into their ASCII values. After that, the Paillier cryptosystem is adopted to encrypt all ASCII values of the original text and send it to the data hider for further processing. At the data hiding phase, the secret data are embedded into homomorphically encrypted text using a technique that does not lose any information, i.e., the homomorphic properties of the Paillier cryptosystem. Finally, the embedded secret data and the original text are recovered at the receiving end without any loss. Experimental results show that the proposed scheme is vital in the context of encrypted text processing at cloud-based services. Moreover, the scheme works well, especially for the embedding phase, text recovery, and performance on different security key sizes.

## I. INTRODUCTION

Privacy and security of personal data are becoming more important as mobile internet and cloud storage technologies continue to advance. Users' personal information may be accessed without their consent by the cloud provider or criminal attackers. Prior to outsourcing, encrypting the data is the most common method for ensuring data confidentiality. In order to identify tampering or declare ownership, a cloud service provider or database management company may be required to insert additional messages, such as authentication or notation data, directly into encrypted cover media. As an example, text-based information may be incorporated into his or her encrypted text to prevent it from being exposed.

To meet the demands of third-party computing platforms (e.g. cloud computing), researchers are increasingly focusing on how to keep data hidden in encrypted domains. Many schemes to conceal data in encrypted images or videos have been reported in the literature in the last few years [1], [2], [3], [4]. As a result of data embedding operations, these techniques deform the cover medium, rendering it unusable for data extraction. Permanent distortion is expressly prohibited in certain delicate situations. This means a legitimate receiver should retrieve the original cover material without mistake following image decryption and data extraction. The RDH in the encrypted domain is chosen to tackle this issue. Basically, the RDH slightly modifies digital material (e.g., image or video) to embed secret data, and the original media may be retrieved when the hidden messages are removed [5]. This data masking approach is beneficial in military communication, medical research, law-enforcement, and error concealment. Lossless compression [6], histogram modification [7], and difference expansion [8], [9], [10] have been developed for RDH.

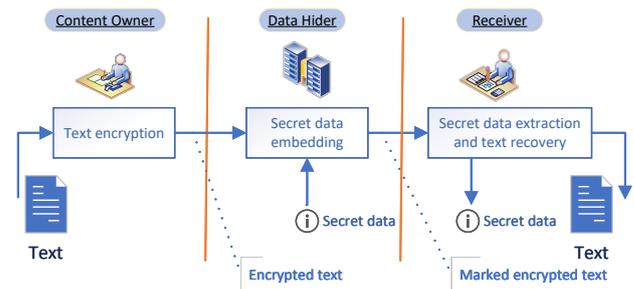

Fig. 1. An example of RDH for encrypted text processing.

Despite the fact that RDH approaches have received a significant amount of attention, RDH in encrypted domains has just emerged as a fresh and difficult area of research. During the course of the last several years, a number of RDH techniques for encrypted image have been suggested. Moreover, many other researchers also adopted the study in RDHED[11], [12] to realize the secret sharing techniques for multiple users in the encrypted domain. In general, these methods can be broken down into two distinct categories: those that involve Vacating Room After Encryption (VRAE) [5], [13], [14], those that involve Reserving Room Before Encryption (RRBE). In VRAE, the content owner encrypts the original signal, then the data hider modifies encrypted data to incorporate extra bits. This framework has easy and efficient end-user operation. As an image's entropy increases, its embedding capability decreases. Data extraction accuracy and image quality are also poor. The RRBE builds a reserving room before encryption in the plaintext domain. This framework's merits include huge embedding capacity and pure reversibility.

This framework may be impractical since it requires content preprocess before encryption. The content owner usually sends simply an encrypted media to the data hider. Many distinct techniques for compressing a marked, encrypted image have been explored on a different front. These approaches use one form of encryption followed by compression systems that can compress an encrypted image using an image coding standard effectively. Puteaux et al.[15] suggested a high-capacity RDH-EI approach that employs prediction and replacement using the most significant bits (MSBs) with an embedding capacity of 0.960 bpp. Puyang et al.[16] suggested an improved RDHEI approach based on a similar concept. They label each pixel value in the original image with two MSBs, resulting in an embedding capacity increase of up to 1.468 bpp. In the method of Yi et al.[17], a small number of pixels are used as reference values to calculate the prediction error of the majority of pixels, and a parametric binary tree labelling method is proposed to differentiate all prediction-errors in order to achieve the maximum embedding capacity of 2.457 bpp. In addition, a high-capacity and separate RDHEI approach with Huffman coding labelling [18] was presented to attain the maximum embedding capacity of 3.75 bpp.

Recently, homomorphic encryption was added to VRAE and RRBE. Chen et al.[19] were the first to propose an RDH technique based on homomorphic encryption using Paillier's additive homomorphic characteristic. It has been shown that the notion of difference expansion may be used to homomorphic encryption, which was first proposed in [19] and then expanded upon by Shiu et al.[20]. Homomorphic encoded RDH has also been studied in this context ([21], [22], [23], [24], [25]). However, following image encryption, public-key cryptosystems like Paillier's lead to data growth. To improve the RDH in the encrypted domain, the additive homomorphic characteristic of modulo operation is used [26], benefit of using encryption is that it prevents data from becoming larger. Most of the work in RDH schemes can be summarized in [1]. In our knowledge, the proposed scheme is the first scheme of RDH in Encrypted Text (RDHET) using the Paillier cryptosystem. The main contribution is that we have extended the idea of RDHED to RDHET, i.e., the functionality of RDHED can also be used for text.

The rest of this paper is structured as follows. The proposed scheme is introduced in Section II. Section III then presents experimental results and analysis to illustrate the superiority and applicability of the proposed scheme. Finally, in Section IV, we provide a conclusion.

## II. PROPOSED SCHEME

The proposed methodology is depicted in Figure 1. It comprises three parties: the content owner, the data hider, and the receiver. The content owner first encrypts the original text with the public key to create an encrypted text. The data-hider can embed some secret data into the encrypted text without knowing the actual content of the original text with the help of public and data hiding (scramble key) keys. The secret data can be extracted at the receiver's end if the receiver only has the data hiding key. The original text can be completely reconstructed if the receiver only has the private key. It should be noted that in the proposed scheme the text and all special symbols are initially converted to their ASCII values. Furthermore, the Paillier cryptosystem is used to encrypt each ASCII value. Finally, during the recovery phase, all ASCII values are converted into character values.

### A. Encryption

At the content owner side, the original text is transformed into its ASCII values, and then each value is encrypted using the Paillier cryptosystem[27] with a public key. $(N, g)$, where $N = p \times q$ is the multiplication of two large prime numbers and $g \in \mathbb{Z}_{N^2}^*$ which meets the condition $gcd(\frac{g^\lambda \bmod N^2 - 1}{N}, N) = 1$. The $\lambda$ is represented as secret key, which follows the condition $\lambda = lcm(p-1, q-1)$. The $gcd$ stands for Greatest Common Divisor(GCD) and $lcm$ is Least Common Multiple(LCM).

Let the original text be $T$, having size $L$. The ASCII value of each lies in the interval $[0, 255]$ and is symbolized by 8 bits. Using the Equation (1), the content owner of the original text encrypts each ASCII value of $T$.

$$T_c(i) = \mathbb{E}[T(i), r] = g^{T(i)} r^N \bmod N^2 \quad (1)$$

where $T_c(i)$ is the representing ciphertext of each original text value $T(i)$, where $i$ follows the range $1 \leq i \leq L$. The $r$ is randomly selected (such that $r \in \mathbb{Z}_N^*$ ) for each ASCII value, this step for random selection leads to achieve semantic security. Moreover, symbol $\mathbb{E}[\cdot]$ denotes the encryption function. Finally, the encrypted text is denoted as $T_e$, which is used by data hider to insert some secret data $S$.

### B. Data Hiding

After obtaining the encrypted text $T_e$, the data hider is able to effectively embed the additional secret data $S$ into $T_e$ in a manner that does not result in any data being lost. The data hider does not need to be aware of the information of original text in order to embed the secret data $S$. This is accomplished by utilising the homomorphic and probabilistic properties of the Paillier cryptosystem, which allows for a lossless data hiding and the production of the marked encrypted text $T_{me}$. The following subsection will discuss the fundamental concept of lossless secret data embedding.

*1) Lossless Secret Data Embedding:* Because we encrypted the ASCII values of original text using the Paillier cryptosystem (public key), the encrypted values now have an additive homomorphic characteristic inherent in them. The additive homomorphism property can be expressed as the following Equation 2, where $\gamma$ and $\delta$ are two integers and operation ($\oplus$) represents the additive homomorphic operation.

$$\mathbb{D}(\mathbb{E}(\gamma) \oplus \mathbb{E}(\delta)) = \gamma + \delta. \quad (2)$$

A special condition could be possible if $\delta=0$, Equation (2) can be presented as Equation (3):

$$\mathbb{D}(\mathbb{E}(\gamma) \oplus \mathbb{E}(0)) = \gamma. \quad (3)$$

with the help of Equation (3), we can quickly explore that the value acquired by $\mathbb{D}(\mathbb{E}(\gamma) \oplus \mathbb{E}(0))$ is the same as the addition of plaintexts $\gamma$ and 0. As a result, when the additive homomorphic property is applied to the encrypted value $\mathbb{E}(0)$, the ciphertext value changes but the original value, i.e. when $\mathbb{E}(\gamma)$ is decrypted, the original value $\gamma$ is returned. However, $\mathbb{E}(\gamma) \oplus \mathbb{E}(0) \neq \mathbb{E}(\gamma)$ follow the condition. The Algorithm 1 may comprehend the comprehensive embedding process.

---

**Algorithm 1:** The secret data embedding

**Input:** $T_e$, $(N, g)$, $S$
**Output:** $T_{me}$

1 Initialization $k = 1$
2 **for** $i = 1$ *to* $L$ **do**
3   **if** $S_k == 0$ **then**
4     **if** $T_e(i) \bmod 2 == 0$ **then**
5       $T_{me}(i) = T_e(i)$           // $1 \leq i \leq L$
6       $k = k + 1$                    // $1 \leq k \leq L$
7       break
8     **else**
9       $T_e(i) = T_e(i) \times g^0 r^N \bmod N^2$
          // $g \in \mathbb{Z}_{N^2}^*$
10      **if** $T_e(i) \bmod 2 == 0$ **then**
11        $T_{me}(i) = T_e(i)$
12        $k = k + 1$
13        break
14      **else**
15        continue
16      **end**
17    **end**
18  **else**
19    **if** $T_e(i) \bmod 2 == 1$ **then**
20      $k = k + 1$ break
21    **else**
22      $T_e(i) = T_e(i) \times g^0 r^N \bmod N^2$
          // $g \in \mathbb{Z}_{N^2}^*$
23      **if** $T_e(i) \bmod 2 == 1$ **then**
24        $T_{me}(i) = T_e(i)$
25        $k = k + 1$ , break
26      **else**
27        continue
28      **end**
29    **end**
30  **end**
31 **end**

---

### C. Secret Data Extraction and Text Recovery

At the receiver end, there are two condition, secret data extraction and text recovery. The secret data can be extracted by the Equation (4) that was embedded at the data hider side, by using the shared scrambling key $\tau$. Notably, the secret data was first encrypted and then embedded into encrypted text using the scrambling key $\tau$ by the data hider. After extraction of secret data, scrambling key $\tau$ is applied to reconstruct the original secret data.

$$S_k = \begin{cases} 0, & \text{if } T_{me}(i) \bmod 2 = 0 \\ 1, & \text{otherwise} \end{cases} \quad (4)$$

The original text is losslessly restored in both conditions, i.e. from marked encrypted text or after secret data extraction only using secret key $(\lambda)$. The Equation (5) shows the decryption process to recover the original text.

$$R(i) = \mathbb{D}[T_{me}(i)] = \frac{L(T_{me}(i)^\lambda \bmod N^2)}{L(g^\lambda \bmod N^2)} \bmod N \quad (5)$$

where $R(i)$ represents the directly decrypted text value (ASCII value) having the condition $L(x) = (x-1)/N$ and $\mathbb{D}[\cdot]$ denotes the decryption function. After decryption of all the values, the final recovered original text is $R$, which is exactly same as the original text after converting into their character values $T$.

## III. EXPERIMENTAL RESULTS AND ANALYSIS

In order to evaluate the performance of the proposed method, the experimental part was conducted on UNIX platform having CPU: AMD Ryzen9 5950X @ 3.40GHz and 64GB RAM (DDR4-3200) and all the algorithms were implemented in Python3. GMP library was also used to write the code.

*1) Embedding Rate:* The bit per bytes (bpb), or how many bytes values are utilised to carry a single bit of secret information, is used to measure how well the proposed approach. If the proposed method demonstrates an embedding rate of 1 bpb, which means that each value (bytes) of the cover text is accountable for carrying one secret bit of information. The following Equation (6) can be used to calculate the Embedding Rate ($ER$). A higher $ER$ value indicates that the proposed methodology has a higher capacity.

$$ER = \frac{\text{Number of secret bits}}{\text{Number of bytes}}. \quad (6)$$

Two text are chosen to evaluate the performance of the proposed scheme, the maximum embedding rate for each text is 1 bpb (bit per bytes), the result can be analysed in Table I. Moreover, Table I also shows that the proposed scheme is evaluated on different key sizes, which means level of Pailler cryptosystem security. Additionally, the time consumptions for each steps: Key Generation (KeyGen), Encryption (Enc), Embedding (Emb) and Decryption (Dec) are calculated. It is noted that, the performance of proposed scheme constant that is embedding rate is 1. The idea behind is that, we have embedded each secret bit into each ASCII value of text.

*2) Comparison and Discussion:* Additionally, aspects of our suggested approach are contrasted with relevant state-of-the-art techniques [19], [20], [28], [23], [22]. According to Table II, all of the schemes' adopted the encryption methods, which is based on the Paillier cryptosystem [27]. The technique

TABLE I
TEXT EXAMPLES WITH THEIR MAXIMUM EMBEDDING CAPACITY AND TIME UTALIZED FOR EACH STEP

| Key size | Text used in the proposed scheme | Size | $ER$ (bpb) | Time used by each step in second | | | |
|---|---|---|---|---|---|---|---|
| | | | | KeyGen | Enc | Emb | Dec |
| 512 | *Joe waited for the train, but the train was late.* | 49 | 1 | 0.00021333 | 0.00782365 | 0.00017003 | 0.00755078 |
| 1024 | *Joe waited for the train, but the train was late.* | 49 | 1 | 0.00353261 | 0.05169327 | 0.00099690 | 0.05368801 |
| 2048 | *Joe waited for the train, but the train was late.* | 49 | 1 | 0.07993346 | 0.37081310 | 0.00765213 | 0.36430381 |
| 512 | *I looked for Mary and Samantha at the bus station, but they arrived at the station before noon and left on the bus before I arrived.* | 132 | 1 | 0.00040362 | 0.02364980 | 0.00048127 | 0.02142619 |
| 1024 | *I looked for Mary and Samantha at the bus station, but they arrived at the station before noon and left on the bus before I arrived.* | 132 | 1 | 0.00387542 | 0.16950080 | 0.00227195 | 0.15765794 |
| 2048 | *I looked for Mary and Samantha at the bus station, but they arrived at the station before noon and left on the bus before I arrived.* | 132 | 1 | 0.04389070 | 1.03851942 | 0.00752360 | 0.983550081 |

TABLE II
COMPARATIVE ANALYSIS OF THE PROPOSED SCHEME WITH OTHER EXISTING SCHEMES

| Schemes | Separable? | Cover Media used | Extra data expansion | $ER$ (bpp) | Quality (dB) | Encryption technique |
|---|---|---|---|---|---|---|
| Chen *et al.* [19] | No | Image | Yes | $\leq 0.5$ | $\approx 40$ | Paillier cryptosystem |
| Shiu *et al.* [20] | Yes | Image | No | $\leq 0.5$ | $\approx 40$ | Paillier cryptosystem |
| Zhang *et al.* [28] | Yes | Image | No | $\leq 1$ | $+\infty$ | Paillier cryptosystem |
| Khan *et al.* [23] | Yes | Image | Yes | $= 1$ | $+\infty$ | Paillier cryptosystem |
| Malik *et al.*[22] | Yes | Image | No | $= 1$ | $+\infty$ | Paillier cryptosystem |
| Proposed | Yes | Text | No | $= 1$ | Exact text | Paillier cryptosystem |

[19] differs from all other schemes in that it is non-separable manner at the receiver's, meaning that the secret data can only be obtained after the encrypted marked media has been decrypted. In comparison, the proposed schemes is separable. The schemes [22], [23] and the proposed scheme show that the maximum embedding rate is 1 bpb and Quality of the directly decrypted image for schemes [22], [23] is $+\infty$ dB. The proposed scheme shows the perfect recovery of text. Moreover the schemes[19], [20] are 0.5 bpp and 40 dB, respectively. Additionally the schemes [22], [28], [20] does not show any further data expansion, schemes [19], [23] do. It means that our proposed scheme is quite similar to techniques [22], [28], [20]. It is noted that the schemes [19], [20], [28], [23], [22] used cover media as image but the proposed scheme is using text as cover media. Our scheme adopted the functionality of RDH in Encrypted Domain (RDHED) but the scheme excelled for the text based encrypted domain i.e Reversible Data Hiding in Encrypted Text (RDHET). To the best of our knowledge this is the first scheme in RDHET.

## IV. CONCLUSIONS

This paper proposes a novel Reversible Data Hiding in Encrypted Text (RDHET), where the encryption technique is used Paillier cryptosystem. We have used the homomorphic and probabilistic property of the Paillier cryptosystem to embed the secret data in a lossless manner. Initially, the content owner encrypts the ASCII values of text by using the public key. Furthermore, In the data hider embeds the secret data into encrypted text using encryption and scrambled keys. In the recovery phase, using the secret key, the original text can be recovered in a separable manner, and the secret data can be reconstructed losslessly with the help of a scrambled key. The Paillier cryptosystem preserves the privacy of the original text on the cloud server without degradation of security level. In the future, we will improve the embedding capacity and reduce the computational cost by adopting lightweight homomorphic encryption techniques.

ACKNOWLEDGMENT

This research was supported by the JSPS KAKENHI Grant Number 22K19777, JST SICORP Grant Number JP-MJSC20C3, and ROIS NII Open Collaborative Research 2022-22S1402, Japan and CCF-Tencent Rhino-Bird Young Faculty Open Research Fund.